\documentclass[fleqn,usenatbib]{mnras}

\usepackage{newtxtext,newtxmath}
\usepackage{bm}

\usepackage[T1]{fontenc}

\DeclareRobustCommand{\VAN}[3]{#2}
\let\VANthebibliography\thebibliography
\def\thebibliography{\DeclareRobustCommand{\VAN}[3]{##3}\VANthebibliography}

\usepackage[dvipdfmx]{graphicx}
\usepackage{ifpdf}
\ifpdf
  \DeclareGraphicsExtensions{.pdf,.png,.jpg}
\else
  \DeclareGraphicsExtensions{.eps}
\fi
\usepackage{amsmath}	

\title[SR shock-clump interaction]{Property of downstream turbulence driven by the special relativistic shock-clump interaction}

\author[Morikawa et al.]{
Kanji Morikawa,$^{1}$\thanks{E-mail: kanji.m1029@eps.s.u-tokyo.ac.jp}
Yutaka Ohira,$^{1}$
and Takumi Ohmura$^{2,3}$
\\
$^{1}$Department of Earth and Planetary Science, The University of Tokyo, 
7-3-1 Hongo, Bunkyo-ku, Tokyo 113-0033, Japan\\
$^{2}$Institute for Cosmic Ray Research, The University of Tokyo, 
5-1-5 Kashiwanoha, Kashiwa, Chiba 277-8582, Japan\\
$^{3}$Division of Science, National Astronomical Observatory of Japan, 2-21-1 Osawa, Mitaka, Tokyo 181-8588, Japan
}

\date{Accepted XXX. Received YYY; in original form ZZZ}

\pubyear{\the\year{}}

\begin{document}
\label{firstpage}
\pagerange{\pageref{firstpage}--\pageref{lastpage}}
\maketitle

\begin{abstract}
Three-dimensional special relativistic magnetohydrodynamic simulations are performed to investigate properties of the downstream turbulence generated by the interaction between a relativistic shock wave and multiple clumps.
We analyze the properties of the downstream turbulence by performing the Helmholtz decomposition. 
It is shown that, in contrast to the non-relativistic shock case, the amplitude of compressive modes is comparable to that of solenoidal modes for the relativistic shock. 
In addition, many reflected shocks propagate in the downstream region. 
The strength of the compressive mode, the solenoidal mode, the reflected shock waves, and the amplified magnetic field depend on the amplitude of the upstream density fluctuations. 
Our simulation results suggest that the wide distribution of 
the ratio of the magnetic energy to the shock kinetic energy, $\epsilon_B$, in gamma-ray burst afterglows is due to the diversity of the gamma-ray burst environment. 
Furthermore, the inhomogeneity of density around high-energy astrophysical objects affects the spectrum of accelerated particles because the reflected shock and turbulence can inject and accelerate non-thermal particles in the shock downstream region. 
The probability distribution of the downstream quantities, power spectra of turbulence, and vortex generation are also analyzed and discussed in this work. 
\end{abstract}

\begin{keywords}
turbulence -- shock waves -- relativistic processes -- dynamo -- MHD 
\end{keywords}



\section{Introduction}
\label{sec:1}
Relativistic shock waves are expected to be generated in high-energy astrophysical phenomena, for example, gamma-ray bursts (GRBs), active galactic nuclei (AGNs), and pulsar wind nebulae (PWNe).
In weaky magnetized plasmas, the relativistic shock wave is mediated by the Weibel instability \citep{Medvedev1999, Kato2007, Spitkovsky2008}.
This kinetic plasma instability causes magnetic turbulence on a very small spatial scale.
On the other hand, some observational and theoretical studies claimed that there should be magnetic turbulence much larger than the kinetic scale.
Some GRB observations show that the value of $\epsilon_{\rm B}$, the ratio of the magnetic energy to the total energy, is spread over many orders of magnitude, $\sim 10^{-9}$~-~$10^{-1}$ \citep{Santana2014, zhang2024}. 
If the magnetic field turbulence is generated by the Weibel instability, it is not clear why the value of $\epsilon_{\rm B}$ is spread so.
In addition, the large-scale magnetic field is required to explain the polarization \citep{Kuwata2023, kuwata2024} and gamma-ray \citep{Huang2022, derishev2024} observations. 
Magnetic fields can provide some important information about the jet and surrounding media through emissions such as synchrotron radiation. 
Moreover, the strength and typical length scale of magnetic fields decide the maximum energy of particles accelerated in the relativistic shock.
Therefore, we need to understand the generation mechanism and the characteristics of the large-scale magnetic turbulence around the relativistic shock waves.

There are density fluctuations in the interstellar medium and stellar winds \citep{Armstrong1995, Chene2020, Moens2022}. 
When a shock wave propagates into the inhomogeneous region, the shock downstream region must be turbulent by the interaction of the shock wave with the inhomogeneous density (shock-clump interactions).  
The generation of the downstream turbulence is confirmed by many numerical simulations in the case of the non-relativistic or mildly relativistic shocks \citep{Giacalone2007, Inoue2009, Inoue2011, Mizuno2011, Mizuno2014, Hu2022}. 
The downstream turbulence consists mainly of solenoidal modes for non-relativistic shocks \citep{Hu2022}.  
The downstream magnetic field is amplified by the solenoidal turbulence.
The relativistic MHD turbulence in the periodic system has been studied extensively, where solenoidal turbulence is isotropically driven by a random external force \citep{Zrake2012, Zrake2013, Takamoto2016, Takamoto2017}. On the other hand, the turbulence driven by the shock-clump interactions is anisotropically driven \citep{Inoue2011}. 
Furthermore, it is not understood what ratio of incompressibility/compressibility is realized in the downstream region of relativistic shocks.
\citet{Tomita2022} investigated the interaction between a relativistic shock and a single high-density clump using the two-dimensional magnetohydrodynamic (MHD) simulation and showed that the dense clump decelerates rapidly after passing the shock front, due to relativistic effects, before turbulence develops. 
Then, the kinetic energy of the shocked clump is quickly converted to the energy of sound waves (compressive modes).
\citet{Morikawa2024} showed that the interaction between a relativistic shock and multiple clumps can generate downstream turbulence, so that the downstream magnetic field is amplified by the turbulence. 
However, it was not investigated what ratio of incompressibility/compressibility is realized in the downstream region of relativistic shocks.

In this paper, we investigate the properties of the turbulence driven by the interaction of a relativistic shock with multiple clumps performing the Helmholtz decomposition.
This paper is organized as follows.
In $\S\ref{sec:2}$, we give the simulation setup and analysis methods.
In $\S\ref{sec:3}$, we show the simulation results and analyze the properties of the downstream turbulence.
In $\S\ref{sec:4}$, we discuss the free energy of the downstream turbulence and the application for particle acceleration.
Finally, we summarize our work in $\S\ref{sec:5}$.

\section{Methods}
\label{sec:2}
We perform the special relativistic three-dimensional MHD simulation.
In this section, we introduce the relativistic MHD equations, numerical setups in this study, and the Helmholtz decomposition which separates the velocity field into solenoidal and compressive modes.


\subsection{Special relativistic MHD}
The relativistic ideal MHD systems are represented by
\begin{eqnarray}
&\frac{\partial}{\partial t}(\Gamma\rho) + \bm{\nabla}\cdot(\Gamma\rho \bm{v}) ~=~0,&\\
\label{eq:number_flux_cons}
&\frac{\partial}{\partial t}\left( \Gamma^2 (\varepsilon+p) \bm{v} + \frac{c}{4\pi}\bm{E}\times\bm{B} \right) +
\bm{\nabla}\left( pc^2+\frac{c^2}{8\pi}(E^2+B^2)\right)\nonumber\\
&+\bm{\nabla}\cdot\left( \Gamma^2 (\varepsilon+p) \bm{v}\bm{v} - \frac{c^2}{4\pi}\left(\bm{E}\bm{E}+\bm{B}\bm{B}\right) \right) ~=~0,&\\
\label{eq:momentum_flux_cons}
&\frac{\partial}{\partial t}\left( \Gamma^2 (\varepsilon+p)  - p + \frac{1}{8\pi}(E^2+B^2) \right) \nonumber\\
&+\bm{\nabla}\cdot\left( \Gamma^2 (\varepsilon+p) \bm{v} +\frac{c}{4\pi}\bm{E}\times\bm{B} \right)~=~0,\\
\label{eq:energy_flux_cons}
&\frac{\partial \bm{B}}{\partial t} ~=~ \nabla\times(\bm{v}\times\bm{B}),&
\label{eq:induction}
\end{eqnarray}
where $c$ is the speed of light, $\rho$ is proper rest mass density, $p$ is the thermal pressure, $\bm{v}$ is the fluid three-velocity, $\Gamma=1/\sqrt{1-(v/c)^2}$ is the Lorentz factor of the fluid, $\bm{B}$ is the magnetic field, $\bm{E}$ is the electric field, and $\varepsilon$ is the internal energy density. 
The electric field and magnetic field are related by the ideal MHD condition, $\bm{E}=-(\bm{v}/c)\times\bm{B}$. 
The first, second, and third equations mean mass, momentum, and energy conservation, respectively. 
The last equation means the induction equation without any resistivity. 
In addition, the equation of state is used in order to close the equation system, 
\begin{eqnarray}
&p~=~(\hat{\gamma} - 1)(\varepsilon-\rho c^2),
\label{eq:eos}
\end{eqnarray}
where $\hat{\gamma}$ is specific heat ratio.
In this work, we set $\hat{\gamma}=4/3$ because the shock velocity is relativistic.


\subsection{Numerical setup}

We solved the above equations by using the code, SRCANS+ \citep{Matsumoto2019}, where the equations are solved with the Harten–Lax–van Leer method as an approximate Riemann solver, the Monotone Upstream-centered scheme for conservation laws method with the van Leer limiter and
the third-order strong stability preserving Runge-Kutta scheme \citep{Gottlieb1998} are adopted.
In addition, we adopt the equation of state proposed in \citet{Mignone2007} for primitive recovery. 
To keep the condition of $\nabla\cdot\bm{B}=0$, we adopted the 9-wave method \citep{Dedner2002, Mignone2010}. 
The stencils in all the simulations are uniform, and the simulation box size is $(L_x/\Delta,~L_y/\Delta,~L_z/\Delta)=(4000,~600,~600) $, where $L_s$ and $\Delta$ are the box size and grid size in the $s$-directions ($s=x,~y,~z$).
The time step is determined to satisfy the Courant–Friedrichs–Lewy (CFL) condition, 
\begin{equation}
\begin{split}
\Delta t ~&=~ \nu \min\left(
\frac{\Delta}{\max\left(|\lambda_{f,+}^x|,~|\lambda_{f,-}^x|\right)},~
\frac{\Delta}{\max\left(|\lambda_{f,+}^y|,~|\lambda_{f,-}^y|\right)}, \right. \\
&~~~~~~~~~~~~~~~~~~~~~~~\left. \frac{\Delta}{\max\left(|\lambda_{f,+}^z|,~|\lambda_{f,-}^z|\right)}
\right),
\end{split}
\label{eq:CFL_condition}
\end{equation}
where $\nu$ is the Courant number and fixed at $\nu=0.1$ in this simulation.
$\lambda_{f,+}^s$ and $\lambda_{f,-}^s$ are the magnetosonic characteristic velocities in the $s$-direction ($s=x,~y,~z$).


The shock normal direction is set to be the $x$-direction. 
The boundary conditions in the $y$- and $z$-directions are periodic. 
The upstream and downstream boundaries in the $x$-direction are injection and escape boundaries, respectively. 
As the initial condition, we prepare the shock structure that satisfies the Rankine-Hugoniot relations in the shock rest frame. 
The quantities in the shock downstream region are calculated using the mean upstream density. 
After the simulation starts, the shock surface fluctuates because of the non-uniform upstream density distribution, but the mean position of the shock surface does not move significantly. 
Thus, the simulation frame is the mean shock rest frame.

As the initial condition in the upstream region, the uniform gas pressure, $p$, and the uniform magnetic field, $B_{\rm up}$, parallel to the $y$-direction in the simulation frame are imposed. 
The magnetic field strength and the gas pressure are given by two parameters, $\sigma=(B_{\rm up}/\Gamma_{\rm sh})^2 / (4\pi \rho_{\rm 0} c^2)$ and $\beta=8\pi p/(B_{\rm up}/\Gamma_{\rm sh})^2$, respectively. 
In this work, we set $\sigma=3.7\times 10^{-6}$ to see the strong magnetic field amplification by the downstream turbulence, and $\beta = 5.4\times10^3$ to avoid the numerical instability at the shock front. 
These values are larger than typical values in the interstellar medium, but sufficiently small to investigate the strong magnetic field turbulence in the downstream region of the strong relativistic shock. 
The upstream Lorentz factor in the simulation frame, $\Gamma_{\rm sh}\approx 5.2$, which corresponds to the shock velocity in the upstream rest frame, is fixed in this simulation.

The initial upstream density distribution is given by
\begin{eqnarray}
\rho_{\rm up}(x,y,z) = \rho_{\rm 0} + \sum_{i} \delta \rho_{i}(x,y,z),
\label{eq:density}
\end{eqnarray}
where $\rho_{\rm 0}$ is the constant density and the individual clumps are represented by
\begin{eqnarray}
\delta \rho_{i}=\left\{
\begin{array}{ll}
~0 & (d > 2R_{\rm cl}) \\
~ \left(\rho_{\rm cl} - \rho_{\rm 0}\right)\left\lbrace1 + \cos\left(\frac{\pi d}{2R_{\rm cl}} \right)\right\rbrace  & ( d \leqq 2R_{\rm cl})~,
\end{array}
\right.\\
\label{eq:density_clump}
d \equiv \sqrt{(x-x_{{\rm c}i})^2\Gamma_{\rm sh}^2+(y-y_{{\rm c}i})^2+(z-z_{{\rm c}i})^2},~~~~~&
\label{eq:distance_from_center}
\end{eqnarray}
where $d$, $R_{\rm cl}$, and $\rho_{\rm cl}$ are the distance from the clump center, the half-width of the clumps, and the density at the half width, respectively. 
$x_{{\rm c}i},~y_{{\rm c}i}$, and $z_{{\rm c}i}$ are the central coordinates of the individual clumps, which are random in the upstream rest frame. 
The clump size is set to be $R_{\rm cl}=L_{y}/8$ in this work.
As represented in Eq.~(\ref{eq:distance_from_center}), the shape of the clumps is contracted in our simulations due to the Lorentz contraction.
The number of the clumps is determined by $f=N_{\rm cl} V_{\rm cl}/L_x L_y L_z \approx 0.25$, where $f$ is the volume filling factor of the clumps, 
and $N_{\rm cl}$ and $V_{\rm cl}= 4\pi R_{\rm cl}^3/3\Gamma_{\rm sh}$ are the number of clumps in the simulation box and the volume of each clump, respectively.
In this work, we perform three simulations for $\rho_{\rm cl}/\rho_{\rm 0}= 1.1,~2$, and $10$.


\subsection{Helmholtz decomposition}
In order to investigate the property of the turbulence generated in the shock downstream region, we perform the Helmholtz decomposition (See, e.g. \citet{Hu2022}) to decompose the four-velocity field, $\bm{u}=\bm{v}/\sqrt{1-(v/c)^2}$, into its solenoidal and compressive components 
\begin{eqnarray}
\boldsymbol{u} = \boldsymbol{u}^{\rm uni}+\boldsymbol{u}^{\rm comp} + \boldsymbol{u}^{\rm sole},
\label{eq:helmholtz_hodge_decomposition_real}
\end{eqnarray}
where $\boldsymbol{u}^{\rm uni}, \boldsymbol{u}^{\rm comp}$, and $\boldsymbol{u}^{\rm sole}$ are the components of the uniform, compressive, and solenoidal velocity field. 
The compressive and solenoidal velocity fields satisfy the curl-free and divergence-free conditions, respectively. 
In the Fourier space, these conditions are represented by
\begin{eqnarray}
\boldsymbol{k}\times\tilde{\bm{u}}^{\rm comp} = 0,\\
\boldsymbol{k}\cdot\tilde{\bm{u}}^{\rm sole} = 0,
\label{eq:helmholtz_hodge_decomposition_fourier}
\end{eqnarray}
where the tildes mean the Fourier transformed value.
Thus, we can decompose the velocity into the compressive and solenoidal components by using vector analysis formulations as
\begin{eqnarray}
\tilde{\bm{u}}^{\rm comp} ~&=&~ -(\hat{\bm{k}}\cdot\tilde{\bm{u}})\hat{\bm{k}},\\
\label{eq:compressive}
\tilde{\bm{u}}^{\rm sole} ~&=&~ (\hat{\bm{k}}\times\tilde{\bm{u}})\times\hat{\bm{k}},
\label{eq:solenoidal}
\end{eqnarray}
where $\hat{\bm{k}}=\bm{k}/|\bm{k}|$ means the unit wave vector. 
To perform the Helmholtz decomposition, we first apply a Fourier transform to the four-velocity field. We then perform the above calculations and the inverse Fourier transform of each component.

\section{Results}
\label{sec:3}
\subsection{Helmholtz decomposed velocity and reflected shocks}
\label{sec:result_helmholtz}
\begin{figure*}
\centering
\includegraphics[width=\textwidth]{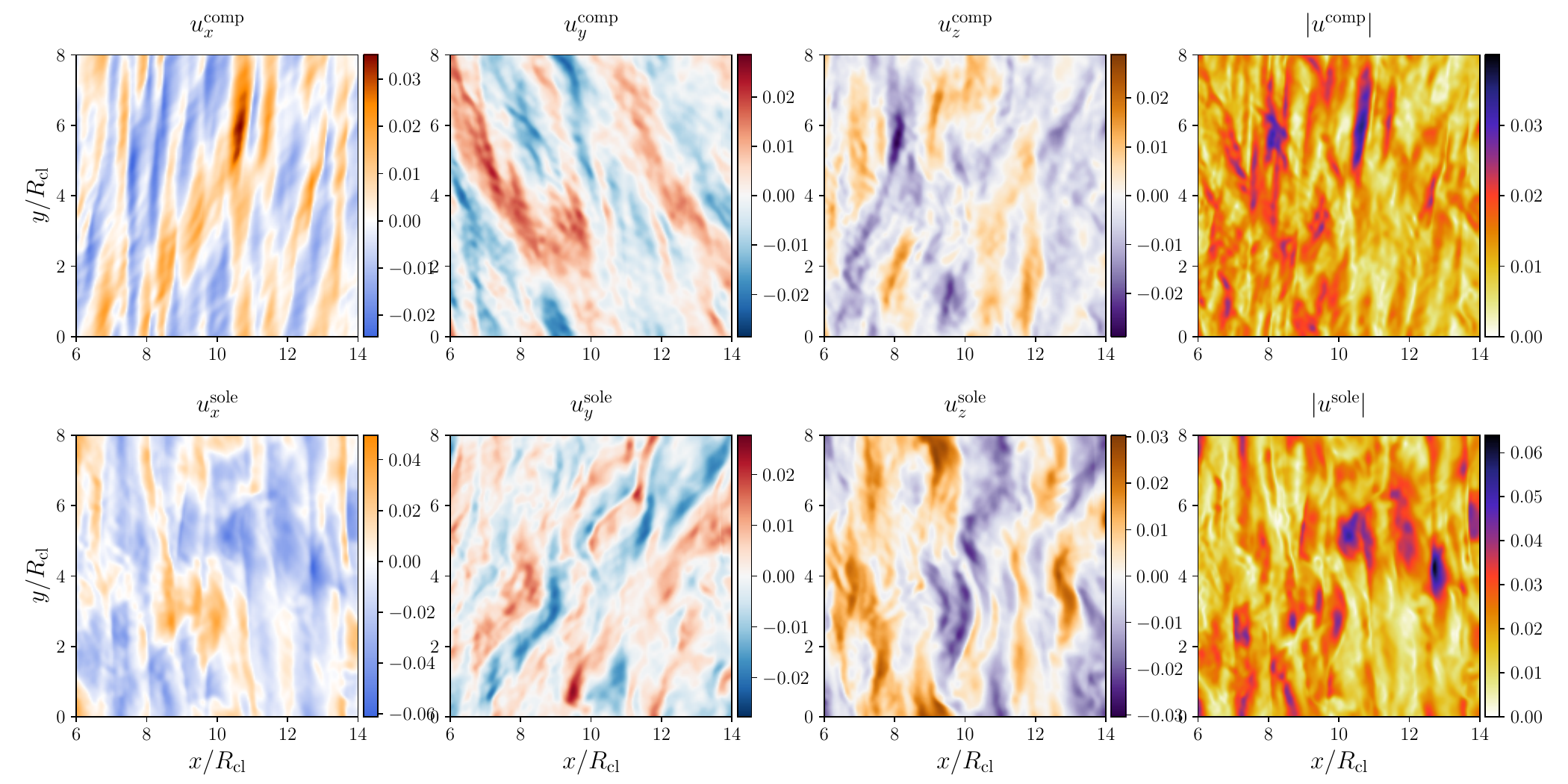}
\caption{
Two-dimensional distributions of $u_x,~u_y,~u_z$, and $|u|$ of the compressive (top four panels) and solenoidal (bottom four panels) modes in the $z=0$ plane at $t\sim56(R_{\rm cl}/c)$ for $\rho_{\rm cl}/\rho_0=1.1$
}
\label{fig:decomposed_delta_rho=1.1_t=210}
\end{figure*}
\begin{figure*}
\centering
\includegraphics[width=\textwidth]{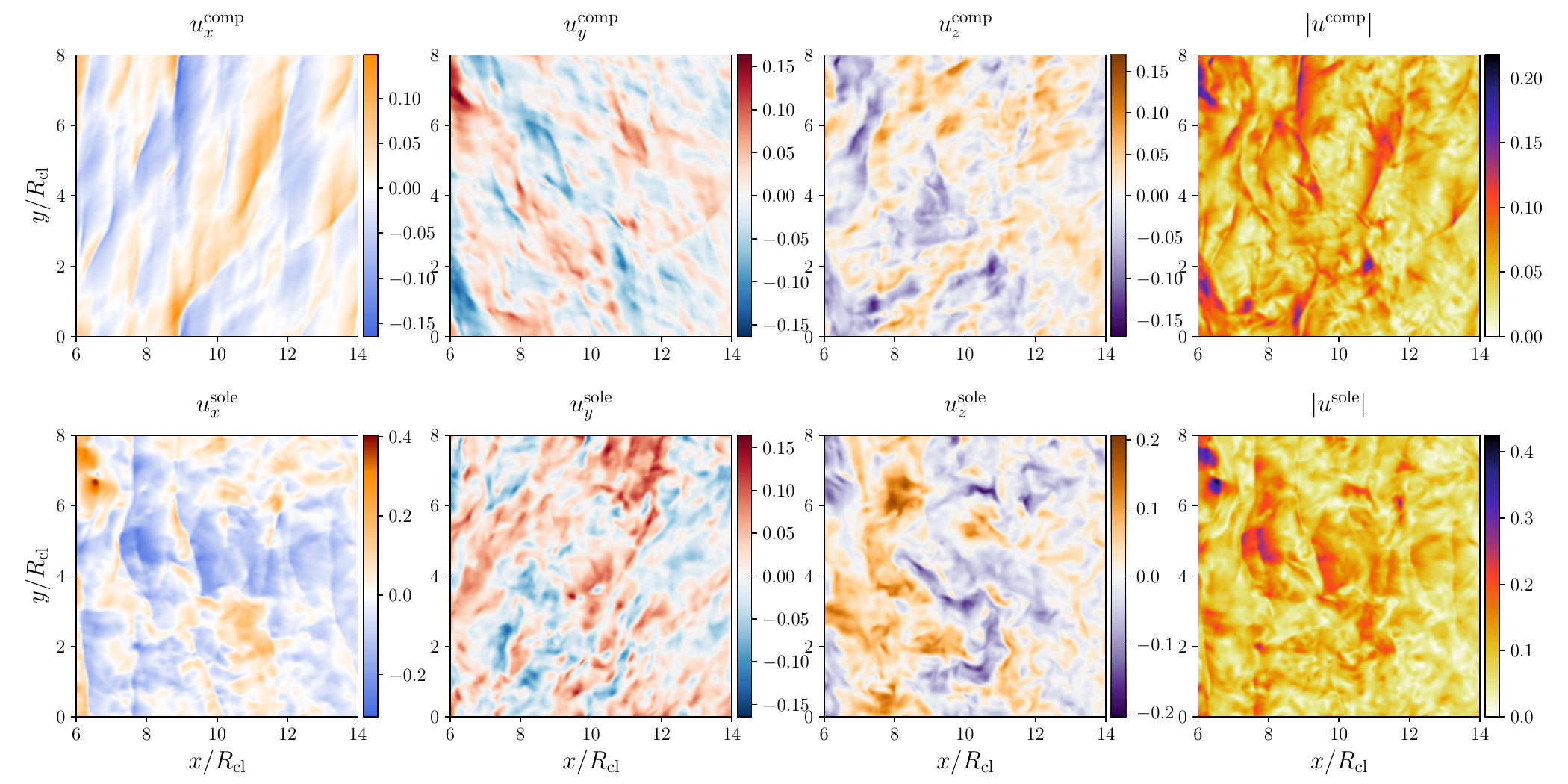}
\caption{
Same as Fig.~\ref{fig:decomposed_delta_rho=1.1_t=210}, but for $\rho_{\rm cl}/\rho_0=2$.
}
\label{fig:decomposed_delta_rho=2_t=210}
\end{figure*}
\begin{figure*}
\centering
\includegraphics[width=\textwidth]{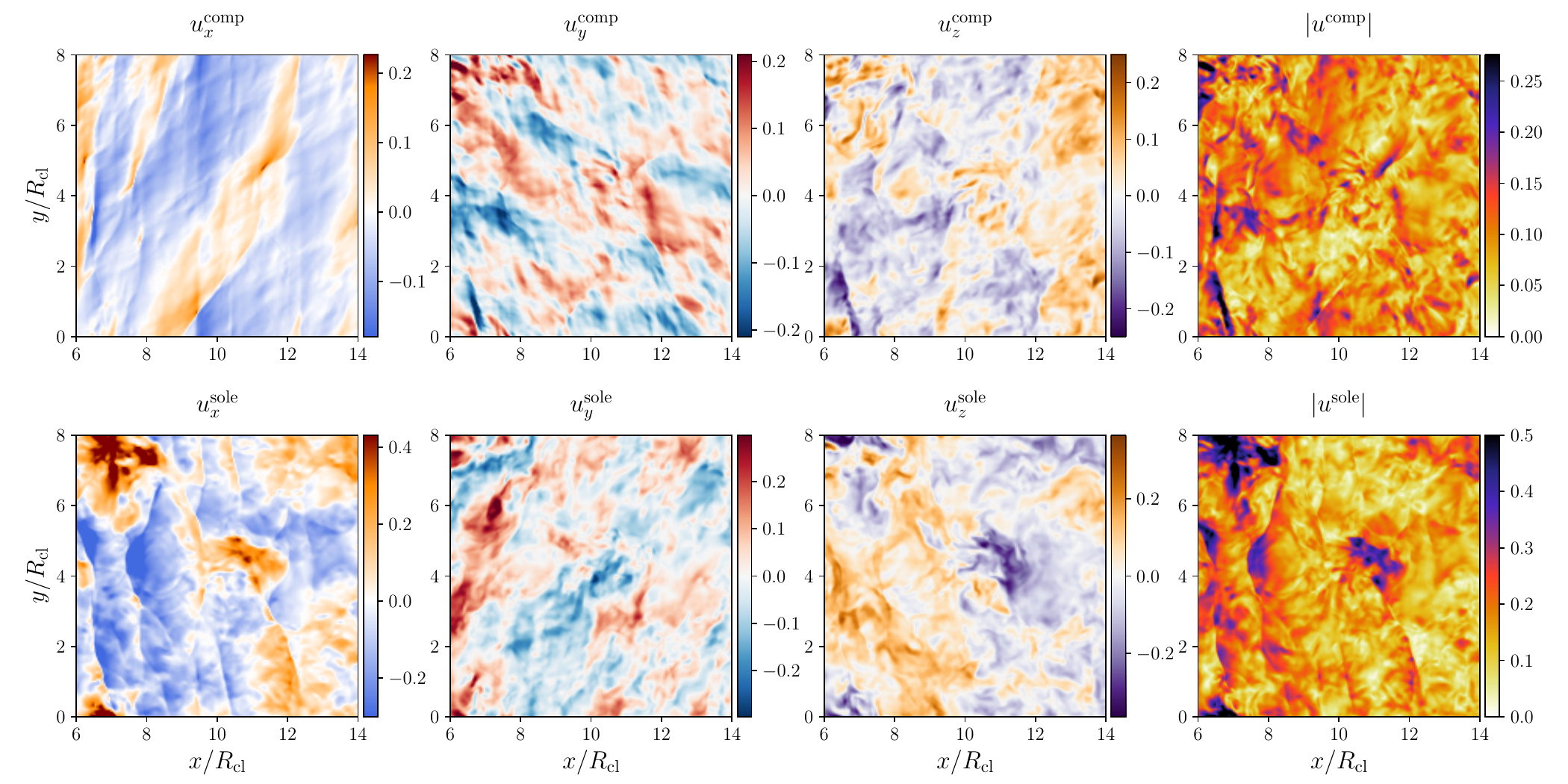}
\caption{
Same as Fig.~\ref{fig:decomposed_delta_rho=1.1_t=210}, but for $\rho_{\rm cl}/\rho_0=10$.
}
\label{fig:decomposed_delta_rho=10_t=210}
\end{figure*}
In our simulation, the main shock front is located at $x/R_{\rm cl}\sim 2.5$, and the region of $x/R_{\rm cl}\gtrsim 2.5$ corresponds to the downstream region, where $R_{\rm cl}$ is the clump size in the upstream rest frame. 
In Figs.~\ref{fig:decomposed_delta_rho=1.1_t=210} - \ref{fig:decomposed_delta_rho=10_t=210}, we plotted two-dimensional distributions of each component of the Helmholtz-decomposed four-velocity field in the $z=0$ plane at $t\sim56(R_{\rm cl}/c)$ for $\rho_{\rm cl}/\rho_0=1.1$, $2$, and $10$. 
The upper and lower rows show velocity fields in the downstream region ($6\leq x/R_{\rm cl}\leq 14$) for compressive and solenoidal modes, where the amplification of the magnetic field by downstream turbulence is almost saturated. 
Although the downstream turbulence is composed mainly of solenoidal modes for the interaction between a non-relativistic shock and clumps \citep{Hu2022}, our simulations show that compressive modes have an amplitude as well as solenoidal modes for relativistic shocks. 
In Tab.~\ref{tab:velocity_components}, we summarize fractions of turbulence in each component, $\langle (u^{s}_{i})^2\rangle/\langle u_{\rm tot}^2\rangle$, where $s={\rm comp}$ and ${\rm sole}$, $i=x,y,$ and $z$, $u_{\rm tot}^2=\sum_{s,i}(u^{s}_{i})^2$, and $\langle X\rangle$ represents the average of $X$ in the region of $6\leq x/R_{\rm cl}\leq 14$.
In all the cases of $\rho_{\rm cl}/\rho_0$, the $x$-component of the solenoidal mode has the largest amplitude, but the compressive mode has an amplitude comparable to that of the solenoidal mode.

\citet{Tomita2022} investigated the interaction between a single clump and the relativistic shock, and showed that the shocked clump quickly decelerates and its kinetic energy is converted to compressive modes if the shock velocity is relativistic ($\Gamma_{\rm sh}\gg \rho_{\rm cl}/\rho_0$). 
In this work, we considered the interaction between multiple clumps and the relativistic shock and showed that the solenoidal mode eventually dominates in the downstream turbulence and the compressive mode also contributes to the downstream turbulence to the comparable levels. 
In addition, the downstream turbulence is not isotropic but has a larger amplitude in the $x$-direction for all cases, which is observed in the non-relativistic case \citep{Inoue2013}. 
This is because the downstream turbulence is driven by the shocked clump with a faster velocity in the $x$-direction.

The panels for $u^{\rm comp}_x$ and $|u^{\rm comp}|$ show some shock-like structures which are generated by the interaction between high-density clumps and the main shock located at $x/R_{\rm cl}\sim2.5$.
On the other hand, the shock-like structure is not seen clearly in the $y$- and $z$-components. 
When a clump collides with the main shock, the downstream plasma of the main shock is pushed in the $x$-direction by the shocked clump due to its large inertia. 
Therefore, the additional shock is generated by the shock-clump interactions. 
These additional shock waves in the downstream region of the main shock are observed in non-relativistic and mildly relativistic shocks \citep{Inoue2010, Inoue2011}.
Hereafter, we call these additional shock waves ``secondary shocks''.

\begin{table}
    \centering
    \begin{tabular}{|c|cccccc|} \hline
       $\rho_{\rm cl}/\rho_0$ & $u_x^{\rm comp}$ & $u_y^{\rm comp}$ & $u_z^{\rm comp}$ & $u_x^{\rm sole}$ & $u_y^{\rm sole}$ & $u_z^{\rm sole}$ \\ \hline
       $1.1$ & 0.15 & 0.14 & 0.12 & 0.28 & 0.13 & 0.18  \\
       $2$ & 0.12 & 0.11 & 0.11 & 0.30 & 0.17 & 0.20  \\ 
       $10$ & 0.13 & 0.13 & 0.13 & 0.27 & 0.16 & 0.18  \\ \hline
    \end{tabular}
    \caption{Fractions of turbulence in each component, $\langle (u^{s}_{i})^2\rangle/\langle u_{\rm tot}^2\rangle$, where $s={\rm comp}$ and ${\rm sole}$, $i=x,y,$ and $z$, $u_{\rm tot}^2=\sum_{s,i}(u^{s}_{i})^2$, and $\langle X\rangle$ means the average of $X$ in the region of $6\leq x/R_{\rm cl}\leq 14$.}
    \label{tab:velocity_components}
\end{table}

\begin{figure*}
    \centering
    \includegraphics[width=\textwidth]{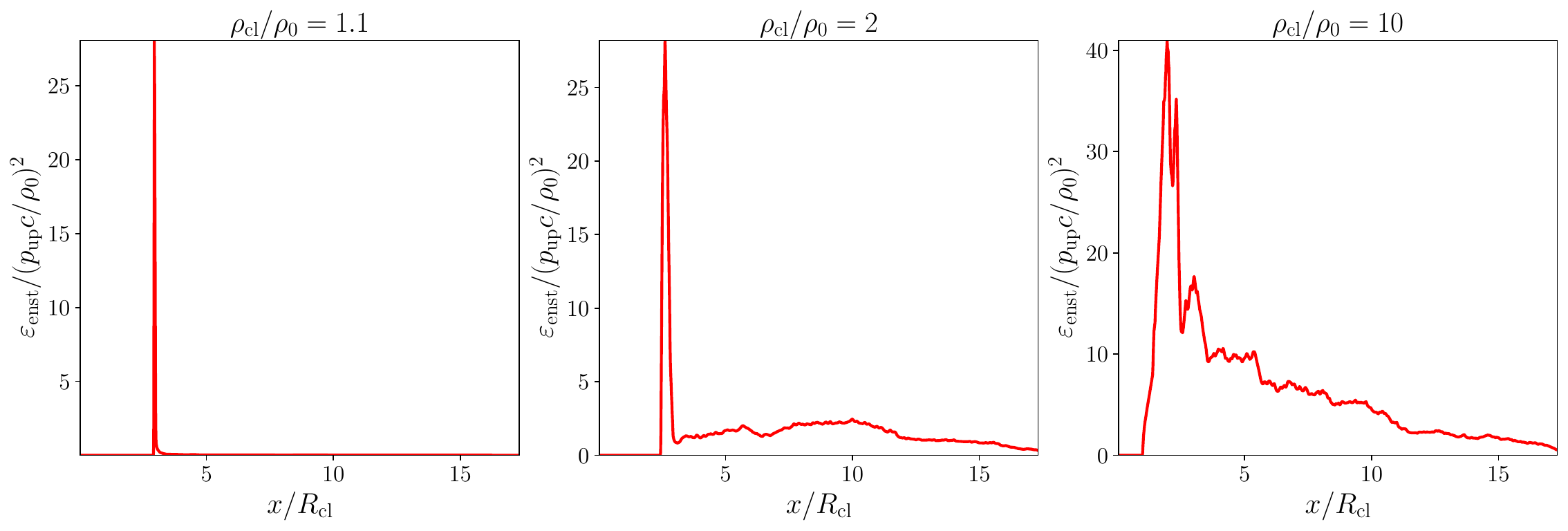}
    \caption{
    The 1D plots in the $x$-direction averaged in the $y$-$z$ plane at $t\sim56(R/{\rm cl}/c)$ are shown here.
    The left panel is for $\rho_{\rm cl}/\rho_0=1.1$, the middle one is for $\rho_{\rm cl}/\rho_0=2$, and the right one is for $\rho_{\rm cl}/\rho_0=10$.
}
    \label{fig:enstrophy}
\end{figure*}

Since the vortex and solenoidal mode can amplify magnetic fields via the turbulent dynamo \citep{Kazantsev1968, Kulsrud1992, Xu2021}, 
we next show the distribution of the vortical energy averaged in the $y$-$z$ plane.
Fig.~\ref{fig:enstrophy} shows one-dimensional plots of the enstrophy density at $t\sim56(R_{\rm cl}/c)$.
The enstrophy density is given by
\begin{eqnarray}
\varepsilon_{\rm enst}=\frac{1}{2}(\omega_{23}^2+\omega_{31}^2+\omega_{12}^2),
\label{eq:enstrophy}
\end{eqnarray}
where the four-vorticity tensor, $\omega_{\mu\nu}$, is given by 
\begin{eqnarray}
\omega_{\mu\nu}=\frac{\partial hu_{\nu}}{\partial x^{\mu}}-\frac{\partial hu_{\mu}}{\partial x^{\nu}}.
\label{eq:four-vorticity}
\end{eqnarray}
Here, $h$ is the enthalpy per rest mass density.
The left, middle, and right panels are for $\rho_{\rm cl}/\rho_0=1.1$, $2$, and $10$, respectively.
The enstrophy has a large spike around the main shock front ($x/R_{\rm cl}\sim2.5$). 
In the case of $\rho_{\rm cl}/\rho_0=1.1$, the enstrophy quickly decays to a very small value behind the shock front. 
However, the other cases show that the enstrophy sometimes increases in the far downstream region. 
In the previous picture, the vorticity is driven only near the main shock front and decays monotonically in the downstream region \citep{Sironi2007,Fraschetti2013}. 
Since there are reflected shocks, sound waves, and shocked clumps (entropy modes) in the downstream region of our simulations, the same mechanisms of vortex generation as that at the main shock can work in the downstream region. 
This could contribute significantly to the mechanism for the generation of the solenoidal mode in the downstream region.
More detailed analyses are required to confirm this picture. 
In addition, further long-term development should also be investigated in future work.


\subsection{Magnetic field amplification}
\label{sec:result_magnetic_field}
\begin{figure}
    \centering
    \includegraphics[width=\linewidth]{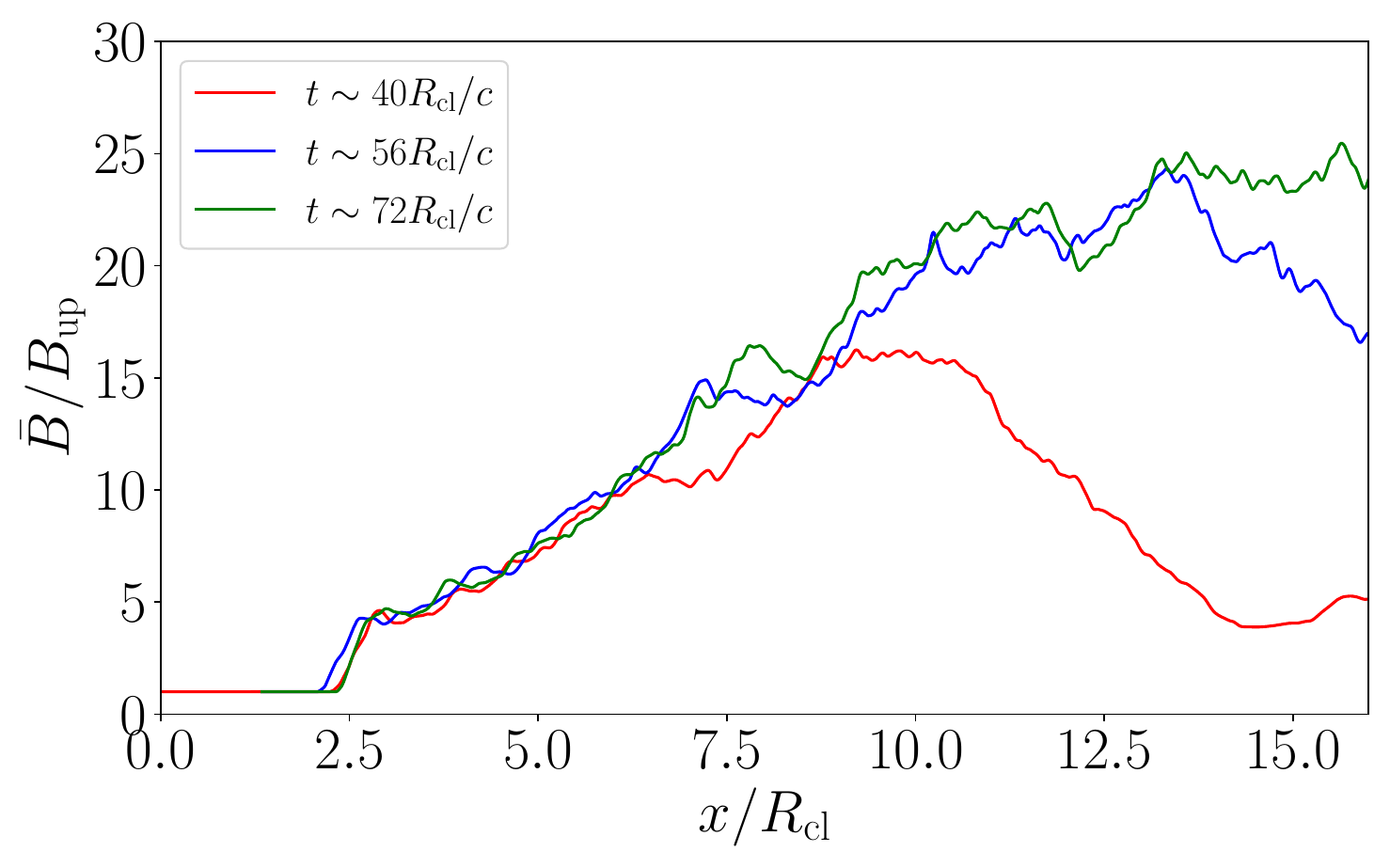}
   \caption{Distribution of the magnetic field strength averaged in the $y$-$z$ plane for $\rho_{\rm cl}/\rho_0=2$. The red blue and green lines show the results for $t \sim 40 R_{\rm cl}/c$, $56 R_{\rm cl}/c$ and $72 R_{\rm cl}/c$. The shock front is located at $x/R_{\rm cl}\sim 2.5$.}
    \label{fig:magnetic_field_time_evolution_delta_rho=1}
\end{figure}
\begin{figure}
    \centering
    \includegraphics[width=\linewidth]{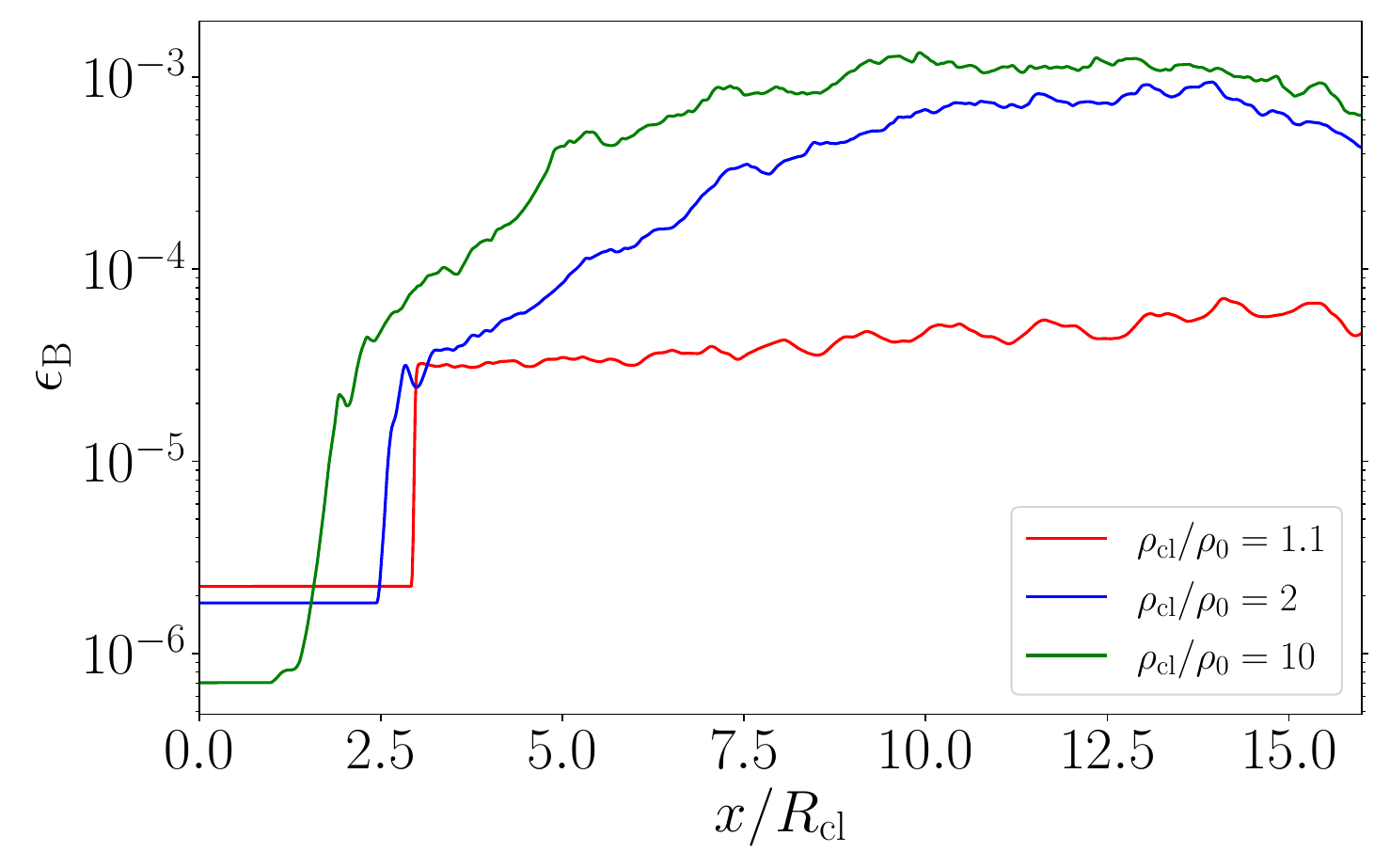}
   \caption{Distribution of the magnetic field energy normalized by the upstream kinetic energy, $\epsilon_{\rm B}$, at $t\sim56 R_{\rm cl}/c$, where the magnetic field energy is averaged in the $y$-$z$ plane. The red, blue, and green lines show $\epsilon_{\rm B}$ for $\rho_{\rm cl}/\rho_0=1.1$, $2$, and $10$, respectively.
   }
    \label{fig:epsilon_b}
\end{figure}
\begin{figure}
    \centering
    \includegraphics[width=\linewidth]{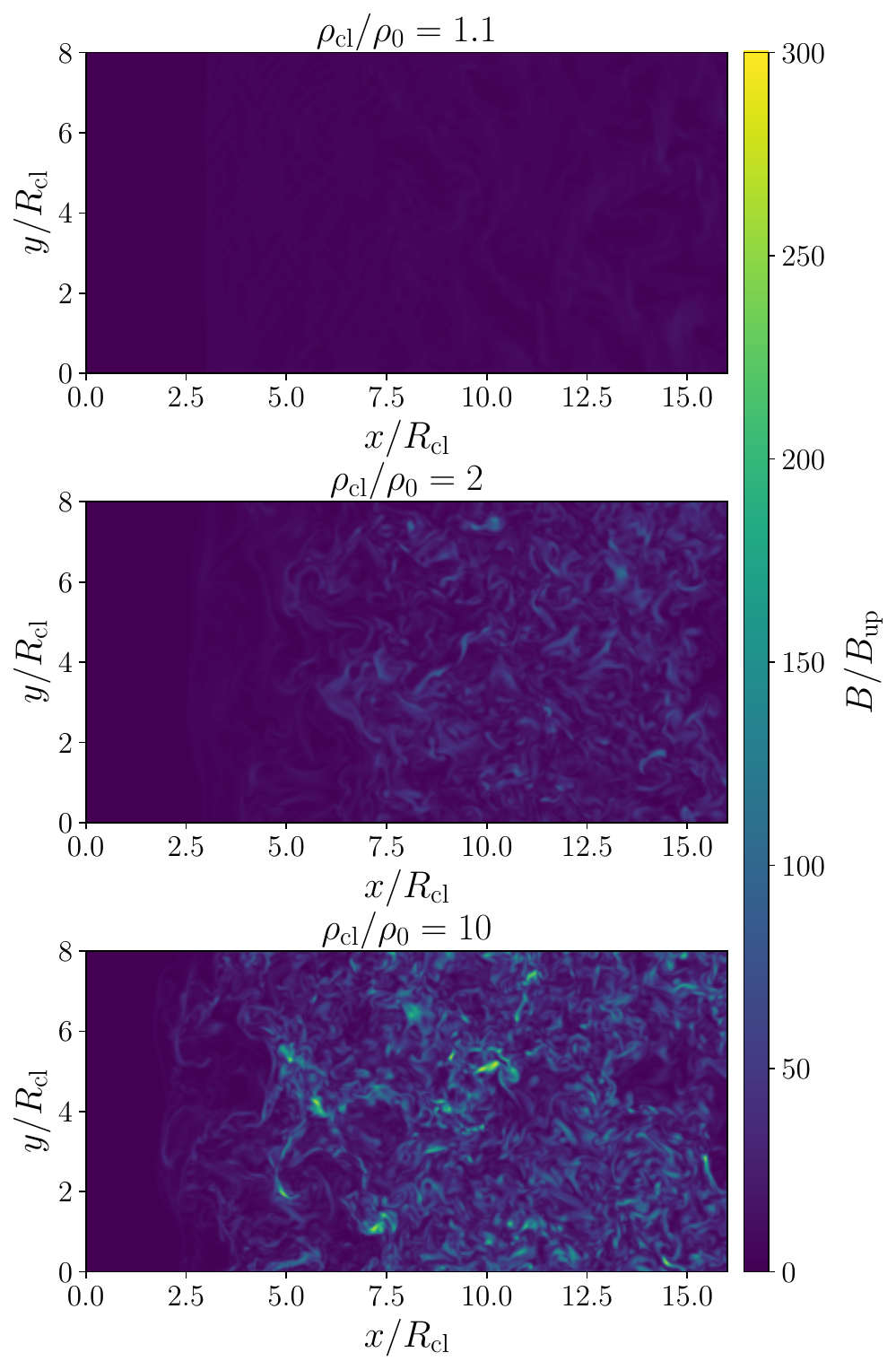}
   \caption{Two-dimensional distribution of the magnetic field strength in the $z=0$ plane at $t\sim56 R_{\rm cl}/c$. The top, middle, and bottom panels show the magnetic field strength for $\rho_{\rm cl}/\rho_0=1.1$, $2$, and $10$, respectively. 
   The shock fronts are located at $x/R_{\rm cl}\sim 2.5$.
   }
    \label{fig:compare_b_1d}
\end{figure}
Fig.~\ref{fig:magnetic_field_time_evolution_delta_rho=1} shows the magnetic field strength averaged in the $y$-$z$ plane for $\rho_{\rm cl}/\rho_0=2$.
The shock front is at $x/R_{\rm cl}\sim2.5$ and the left region to the shock front is the upstream region. 
At first, the magnetic field is amplified by shock compression just behind the shock front.
The amplitude of the shock compression is consistent with the expected value.
The magnetic field amplification in the downstream region is saturated at the time of $t\sim56(R_{\rm cl}/c)$.
The magnetic field is saturated from around $10R_{\rm cl}$ behind the shock front.
This means that it takes around $30 R_{\rm cl}/c$ to be saturated because the downstream flow velocity is $c/3$ in the shock rest frame.
As shown in Fig.~\ref{fig:decomposed_delta_rho=2_t=210}, the solenoidal velocity of turbulence is about $0.1c$.
Thus, the saturation time of the magnetic field is comparable to the eddy turnover time of the turbulence.

To see the dependence of the simulation results on the clump density, the ratios of the magnetic energy to the upstream kinetic energy, $\epsilon_{\rm B}=B^2/8\pi\bar{\rho}_{\rm up}(\Gamma_{\rm sh}-1)c^2$, at $t\sim56(R_{\rm cl}/c)$ for $\rho_{\rm cl}/\rho_0=1.1~,2~,$ and $10$ are shown in Fig.~\ref{fig:epsilon_b}, where $\bar{\rho}_{\rm up}$ is the mean upstream density.
The magnetic energy is amplified by shock compression around $x/R_{\rm cl}\sim2.5$ for all cases.
The magnetic field is strongly amplified in the downstream region for $\rho_{\rm cl}/\rho_0=2$ and $10$, but not for $\rho_{\rm cl}/\rho_0=1.1$. 
From these results, it might be possible to say that the wide distribution of $\epsilon_{\rm B}$ in the observation of GRB afterglows indicates the diversity of the GRB environment. 
This should now be kept in mind when trying to understand the origins of GRBs. 
$\epsilon_{\rm B}$ is $3$~-~$4 \times 10^{-4}$ for $\rho_{\rm cl}/\rho_0=2$ and $10$ in our previous study with lower resolution \citep{Morikawa2024}, but $\epsilon_{\rm B}\approx 7\times 10^{-4}$ and $10^{-3}$ for $\rho_{\rm cl}/\rho_0=2$ and $10$ in this study with higher resolution. 
Although a lower resolution of the MHD simulation weakens the magnetic field amplification by turbulence \citep{Ji2016}, $\epsilon_{\rm B}$ will definitely be at least around $10^{-4}$~-~$10^{-3}$ for $\rho_{\rm cl}/\rho_0=2$ and $10$. 
The amplified magnetic field may be stronger for higher resolutions.

At Eq.~(36) in \citet{Sironi2007}, $\epsilon_{\rm B}$ is estimated as
\begin{eqnarray}
\epsilon_{\rm B}~\approx~3.6\times10^{-2}\left(\frac{\Gamma_{\rm sh}}{5}\right)^{-1}\left(\frac{\rho_{\rm cl}}{2}\right)^{2}\left(\frac{f}{0.25}\right),
\label{eq:epsilon_b_sg2007}
\end{eqnarray}
where the energy of the vortex generated by the shock-clump interaction is assumed to be converted to the energy of magnetic fields.
This value is larger than those in our simulations. 
In \citet{Sironi2007}, they estimated the energy of the vortex from the velocity of the shocked clump just behind (downstream side) the shock front. However, as shown in \citet{Tomita2022}, the shocked clump is quickly decelerated in the downstream region, so that the kinetic energy of the shocked clump is converted to the energy of sound waves. 
As shown in Fig. \ref{fig:decomposed_delta_rho=10_t=210}, these sound waves are dissipated by the reflected shock. 
Thus, the kinetic energy of the shocked clumps is converted to thermal energy before it is converted to magnetic field energy, resulting in a smaller $\epsilon_{\rm B}$ than that in Eq.~(\ref{eq:epsilon_b_sg2007}).

Fig.~\ref{fig:compare_b_1d} shows the distribution of the magnetic field strength on the $z=0$ plane at $t\sim56(R_{\rm cl}/c)$ for $\rho_{\rm cl}/\rho_0=1.1,~2,~$ and $10$. 
The magnetic field is amplified more strongly as the density of the clumps increases.  
The maximum magnetic field strength is about $15B_{\rm up}$, $100B_{\rm up}$, and $300B_{\rm up}$ for $\rho_{\rm cl}/\rho_0=1.1,~2,~$ and $10$, respectively. 
As shown in Fig.~\ref{fig:pdf} (See the details in Section~\ref{sec:PDF}), the distribution of the magnetic field is the log-normal distribution.
The average value is about $5B_{\rm up}$, $25B_{\rm up}$, and $40B_{\rm up}$ for $\rho_{\rm cl}/\rho_0=1.1,~2,$ and $10$, respectively.

For $\rho_{\rm cl}/\rho_0=1.1$ in this work, the magnetic field is not strongly amplified, but $\epsilon_{\rm B}$ is gradually increasing in the downstream region. 
These are due to the small kinetic energy and long eddy turnover time of turbulence. 
If the upstream magnetic field is sufficiently small, the magnetic field is expected to be strongly amplified in the downstream region even for $\rho_{\rm cl}/\rho_0=1.1$, but it will take longer because the eddy turnover time is longer for $\rho_{\rm cl}/\rho_0=1.1$. 
This will be examined in the future.


\subsection{Fourier analysis}
\label{sec:result_fourier_analysis}

\begin{figure*}
    \centering
    \includegraphics[width=\textwidth]{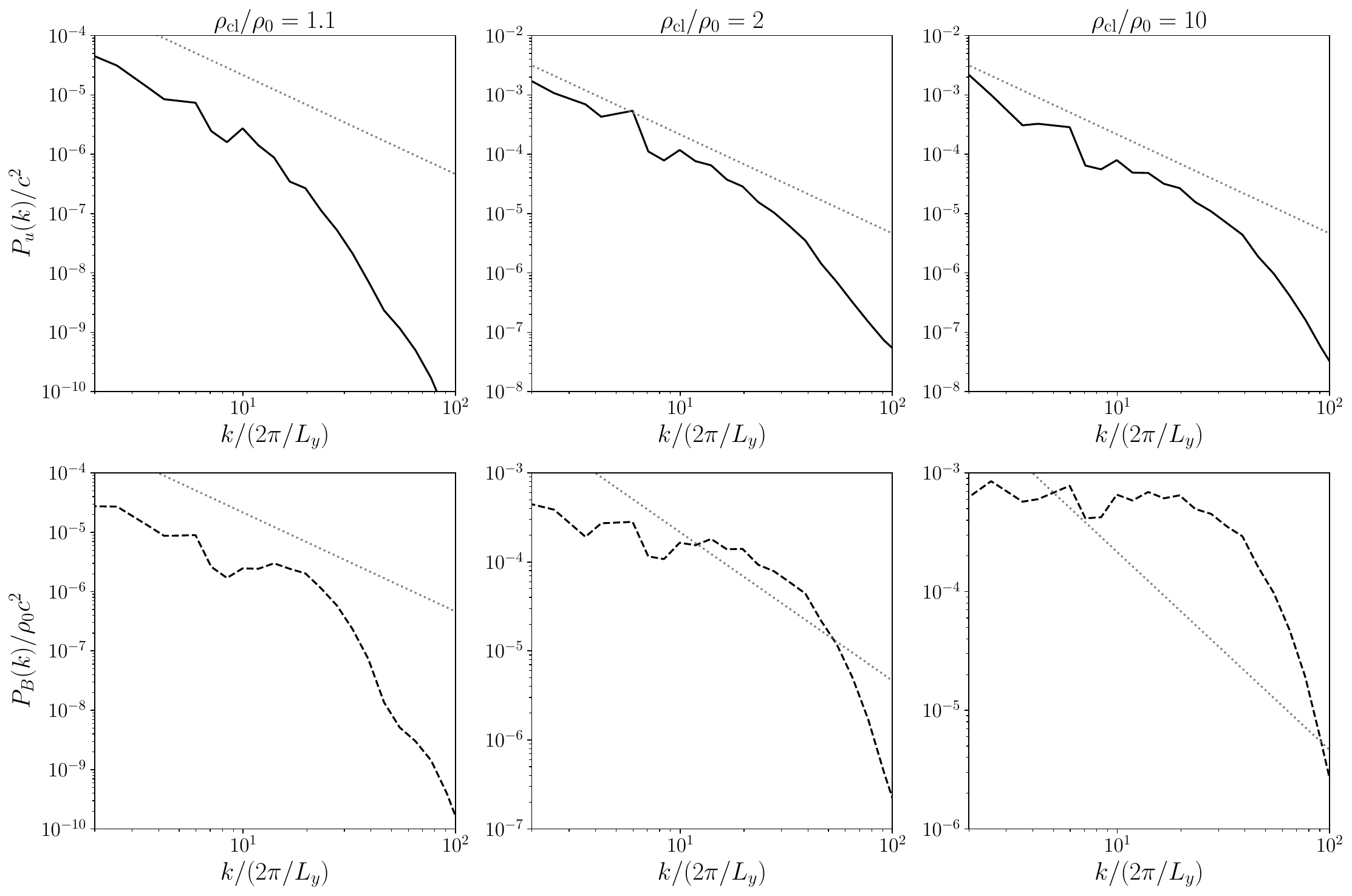}
   \caption{Power spectra of the four-velocity (solid line) and magnetic field (dashed line) in the downstream region ($6\leq x/R_{\rm cl}\leq 14$) at $t/(R_{\rm cl}/c)\sim56$. The left, middle, and right panels are results for $\rho_{\rm cl}/\rho_0=1.1, 2$, and $10$, respectively. The dotted line in each panel is the line proportional to $k^{-5/3}$.}
    \label{fig:fourier_spectrum}
\end{figure*}
\begin{figure*}
    \centering
    \includegraphics[width=\textwidth]{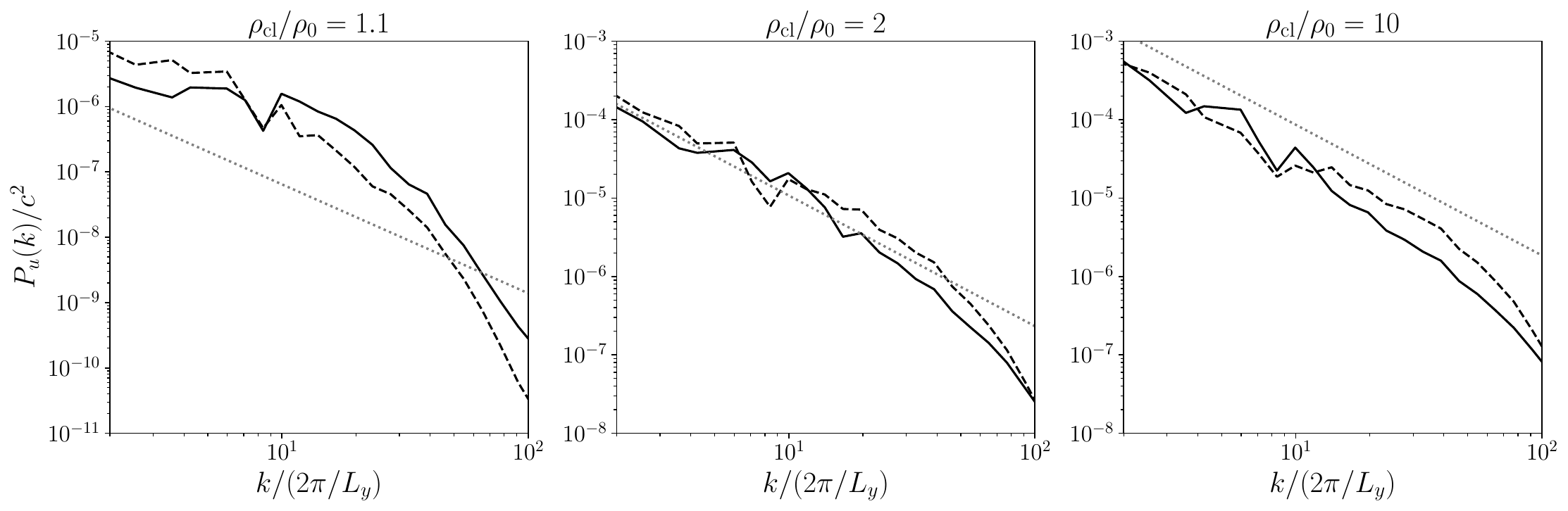}
   \caption{Same as Fig.~\ref{fig:fourier_spectrum}, but for the Helmholtz-decomposed compressive (solid line) and solenoidal (dashed line) velocity fields.}
    \label{fig:fourier_spectrum_decomposed}
\end{figure*}
In this section, we examine the turbulent property in the Fourier space.
After turbulence cascades into the small scale, the scale-independent structure is expected to appear in the so-called inertial range.
Similarly to previous studies, we define the power spectrum of four-velocity as follows.
\begin{eqnarray}
P_u(k)dk~=~\frac{1}{2}\sum_{\bm{k}\in dk}\tilde{\boldsymbol{u}}_{\bm{k}}\cdot\tilde{\boldsymbol{u}}^{*}_{\bm{k}}.
\label{eq:def_energy_spectrum}
\end{eqnarray}
We also define the power spectrum of magnetic fields as
\begin{eqnarray}
P_B(k)dk~=~\frac{1}{8\pi}\sum_{\bm{k}\in dk}\tilde{\boldsymbol{B}}_{\bm{k}}\cdot\tilde{\boldsymbol{B}}^{*}_{\bm{k}}.
\label{eq:def_magnetic_energy_spectrum}
\end{eqnarray}
Since the simulation in this work is not the periodic system in the $x$-direction, 
in order to reduce the spectral distortion due to the non-periodic boundary, the Hamming window function is used to perform the Fourier transform in the $x$-direction.

\begin{figure*}
    \centering
    \includegraphics[width=\textwidth]{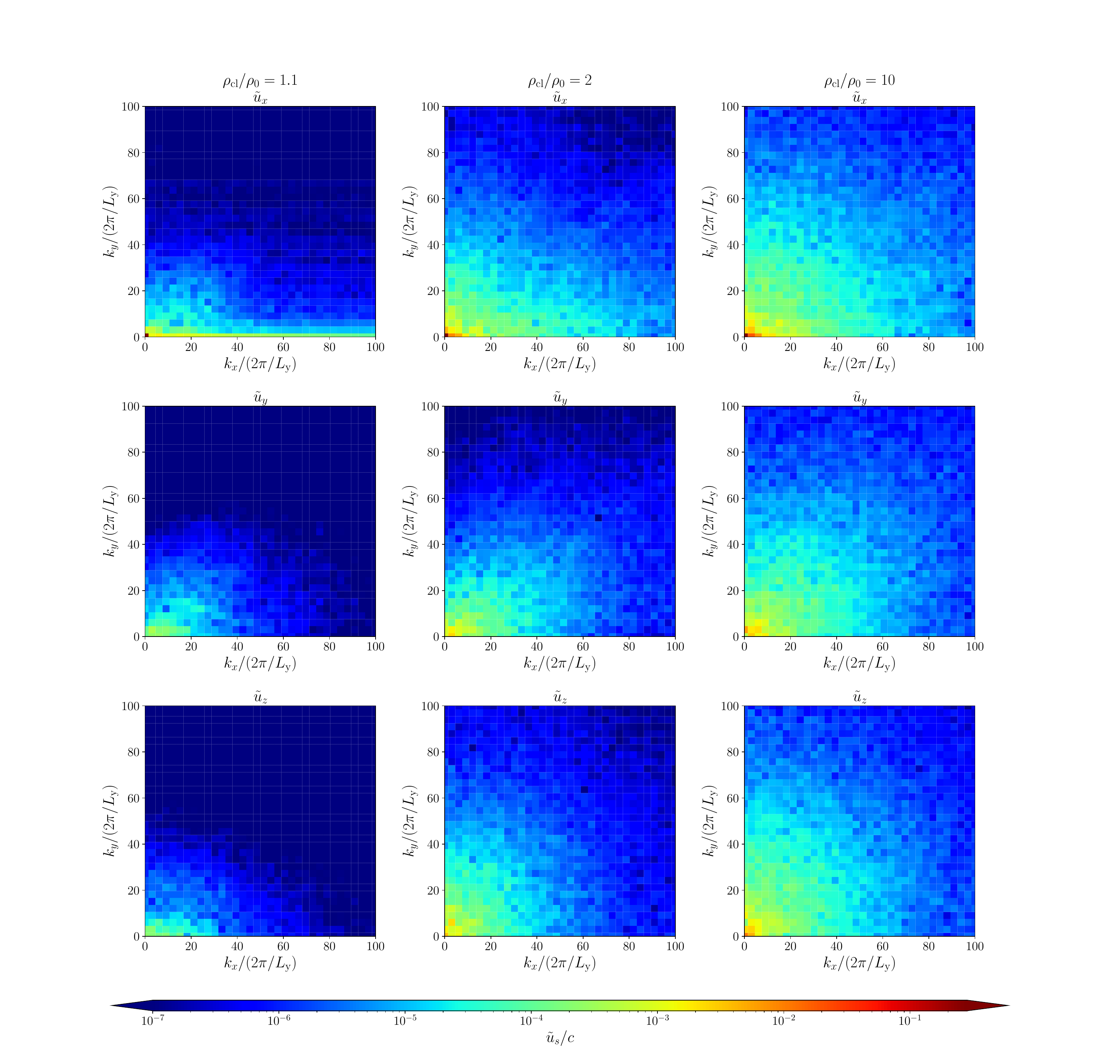}
   \caption{
   Fourier transformed four-velocity in the plane of $k_z=0$, where the velocity fields are taken from the downstream region ($6\leq x/R_{\rm cl}\leq 14$). 
   The upper, middle, and bottom rows show $\tilde{u}_x$, $\tilde{u}_y$, and $\tilde{u}_z$, respectively.
   The left, middle, and right columns are in the case of $\rho_{\rm cl}/\rho_0=1.1$, $2$, and $10$, respectively. 
   }
    \label{fig:fourier_2d}
\end{figure*}
The solid and dashed lines in Fig.~\ref{fig:fourier_spectrum} show power spectra of the four-velocity and magnetic field in the downstream region ($6\leq x/R_{\rm cl}\leq 14$) at $t\sim56(R_{\rm cl}/c)$, respectively. 
The left, middle, and right panels are results for $\rho_{\rm cl}/\rho_0=1.1$, $2$, and $10$, respectively. 
The dotted line in each panel is the line proportional to $k^{-5/3}$ as a reference.
As expected, the amplitude of the power spectra increases with the density of clumps. 
The spectrum of four-velocity has a power-law form of $k^{-5/3}$ in the small $k$ range and a cutoff shape in the large $k$ range for all cases. 
Because this work does not take into account the physical viscosity, the cutoff structure is due to the numerical viscosity. 
The spectra of magnetic fields have flatter spectra than those of the four-velocity for $\rho_{\rm cl}/\rho_0=2$ and $10$.
In the case of $\rho_{\rm cl}/\rho_0=1.1$, the inertial range is very narrow. This is because the turbulence is weaker and the eddy turnover time is longer than that for other cases. Before cascade to the smaller scale, the turbulence is dissipated by the numerical viscosity.

In Fig.~\ref{fig:fourier_spectrum_decomposed}, we show the power spectra for the Helmholtz-decomposed four-velocities.
The solid and dashed lines are the spectra for compressive and solenoidal modes, respectively, and the dotted line in each panel is the line proportional to $k^{-5/3}$.
The left, middle, and right panels show results for $\rho_{\rm cl}/\rho_0=1.1$, $2$, and $10$, respectively. 
For $\rho_{\rm cl}/\rho_0=2$ and $10$, the solenoidal and compressive modes are not so different from each other, but the solenoidal spectrum has a larger amplitude than the compressive one, especially in the large $k$ region. 
On the other hand, for $\rho_{\rm cl}/\rho_0=1.1$, the compressive spectrum has a larger amplitude than the solenoidal one in the large $k$ region.

In terms of the spectral index of the turbulence, in the case of $\rho_{\rm cl}/\rho_0=1.1$, the spectrum of the compressive mode seems to be flatter than $k^{-5/3}$ and that of the solenoidal mode although the power law region is very small. 
The acoustic turbulent theory predicts the spectrum of $k^{-3/2}$ \citep{zakharov1970, Cho2002, Galtier2023, Kochurin2024}.
Larger simulations are required to determine exactly which power-law index the simulation results indicate. 
On the other hand, in the cases of $\rho_{\rm cl}/\rho_0=2$ and $10$, both the spectra of the compressive and solenoidal modes seem to be proportional to $k^{-5/3}$.

To see the anisotropy of the downstream turbulence ($6\leq x/R_{\rm cl}\leq 14$) in the wavenumber space, in Fig.~\ref{fig:fourier_2d}, we show the Fourier transformed four-velocity in the $k_z=0$ plane at $t\sim 56(R_{\rm cl}/c)$. 
The left, middle, and right columns are in the case of $\rho_{\rm cl}/\rho_0=1.1, 2, 10$, and the top, middle, and bottom rows show $u_x$, $u_y$, and $u_z$, respectively. 
As can be seen, the turbulence cascades selectively in the $x$-direction for $\rho_{\rm cl}/\rho_0=1.1$. 
The direction of the cascade is more isotropic as the density of the clump increases.

For $\rho_{\rm cl}/\rho_0=1.1$, the downstream turbulence is mainly composed of the acoustic (or fast) mode in the large $k$ region ($k/(2\pi/L_y) \gtrsim 10$) as shown in Fig.~\ref{fig:fourier_spectrum_decomposed}. 
The dispersion relation of the acoustic mode is represented by
\begin{eqnarray}
\omega(\bm{k})~=~\pm kc_{\rm s},
\label{eq:acoustic_velocity}
\end{eqnarray}
where $\omega$ and $c_{\rm s}$ are the wave frequency and the acoustic velocity, respectively.
Since the downstream plasma beta is very large in this work, the phase velocity of the fast mode is almost the same as the acoustic velocity. 
Following the resonance condition of the three-wave interaction, the acoustic mode cascades as
\begin{eqnarray}
\omega(\bm{k})~&=&~\omega(\bm{k}_1)+\omega(\bm{k}_2),\\
\label{eq:resonant_omega}
\bm{k}~&=&~\bm{k}_1+\bm{k}_2,
\label{eq:resonant_k}
\end{eqnarray}
where $k_1$ and $k_2$ are the colinear wave vector \citep{Kochurin2024}, which is the same picture as the Iroshnikov-Kraichnan turbulence. 
The acoustic mode excited by the shock-clump interaction propagates in the $x$-direction. 
For $\rho_{\rm cl}/\rho_0=1.1$, the nonlinear interaction between the acoustic mode and the shocked clump is weak. 
Then, all acoustic modes propagate mainly in the $x$-direction. 
As a result, the cascade through the three-wave interaction develops in the $x$-direction. 
On the other hand, in the cases of $\rho_{\rm cl}/\rho_0=2$ and $\rho_{\rm cl}/\rho_0=10$, the shock front is more distorted, and the nonlinear interaction in the downstream region is expected to be stronger than that for $\rho_{\rm cl}/\rho_0=1.1$. 
Then, many modes would be more isotropically generated, so that the cascade would be more isotropic. 


\subsection{Probability density function}
\label{sec:PDF}

\begin{figure*}
    \centering
    \includegraphics[width=\textwidth]{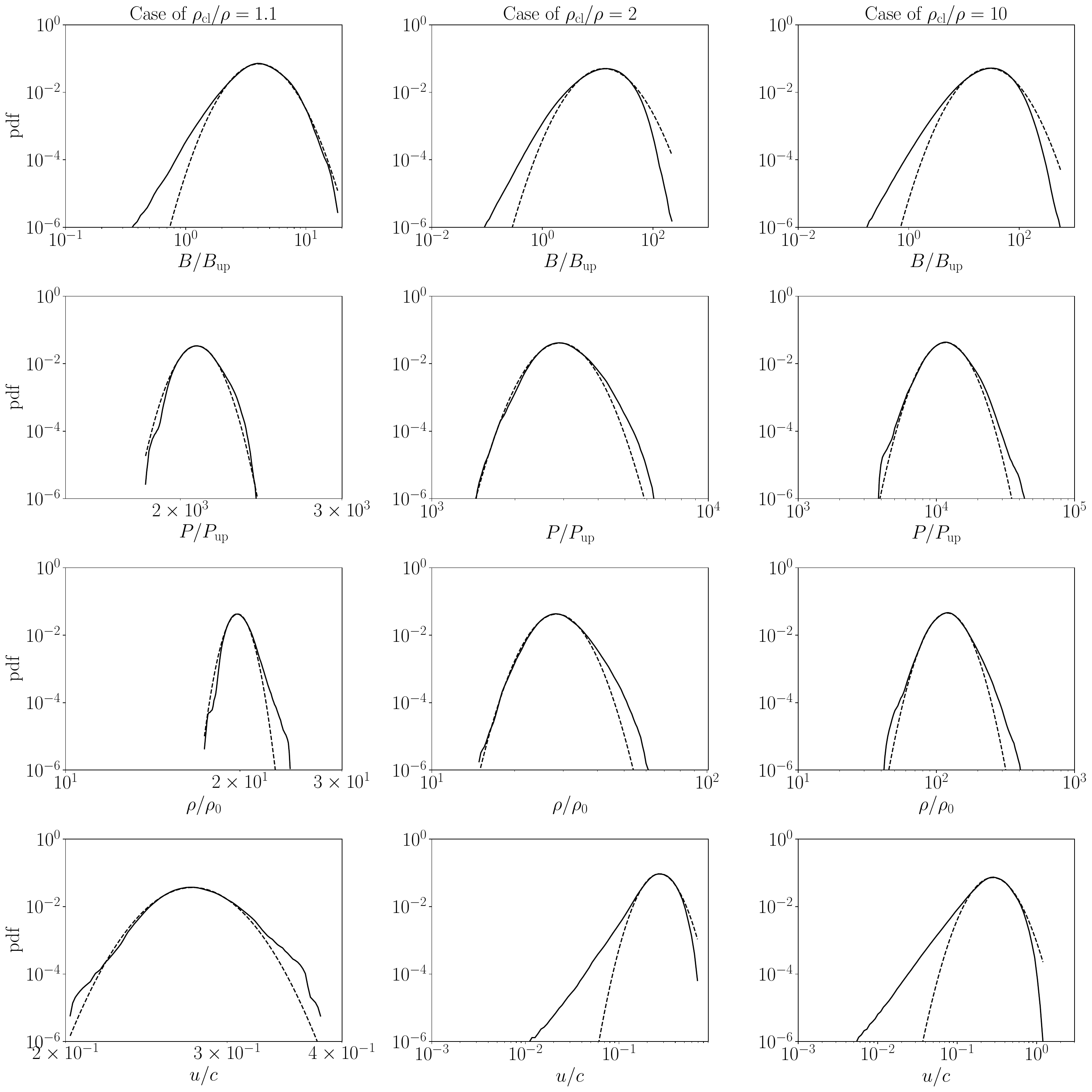}
   \caption{Probability distribution for the magnetic field strength, pressure, proper density, and the absolute value of the four-velocity in the downstream region ($6\leq x/R_{\rm cl}\leq 14$). 
   The solid lines are the probability distribution in the simulation and the dashed lines are log-normal functions fitted to the corresponding simulation data.
   The results for $\rho_{\rm cl}/\rho_0=1.1$, $2$, and $10$ are shown in order from left to right columns.}
    \label{fig:pdf}
\end{figure*}
The probability distribution of the physical value in the downstream region ($6\leq x/R_{\rm cl}\leq 14$) is shown in Fig.~\ref{fig:pdf}. 
The left, middle, and right columns show the results for $\rho_{\rm cl}/\rho_0=1.1$, $2$, and $10$, respectively. 
The probability distributions of the magnetic field strength, pressure, density, and absolute value of the four-velocity are plotted in order from top to bottom rows. 
The solid lines show the simulation results, and the dashed lines are log-normal functions fitted to the simulation data. 
The simulation results of the density and pressure are described roughly by the log-normal distribution, but large deviations from the log-normal distribution are observed in other quantities. 
For the magnetic field strength and four-velocity, significant deviations appear in the small values, which are observed in \citet{Inoue2011}. 
They showed that the probability distribution of the magnetic field can be fitted by $B^2\exp{(-B/a(t))}$, where $a(t)$ is a function of time. 
The magnetic reconnection region might be the origin of the low magnetic field region \citep{Lazarian1999, Xu2016}.

\begin{figure*}
    \centering
    \includegraphics[width=\textwidth]{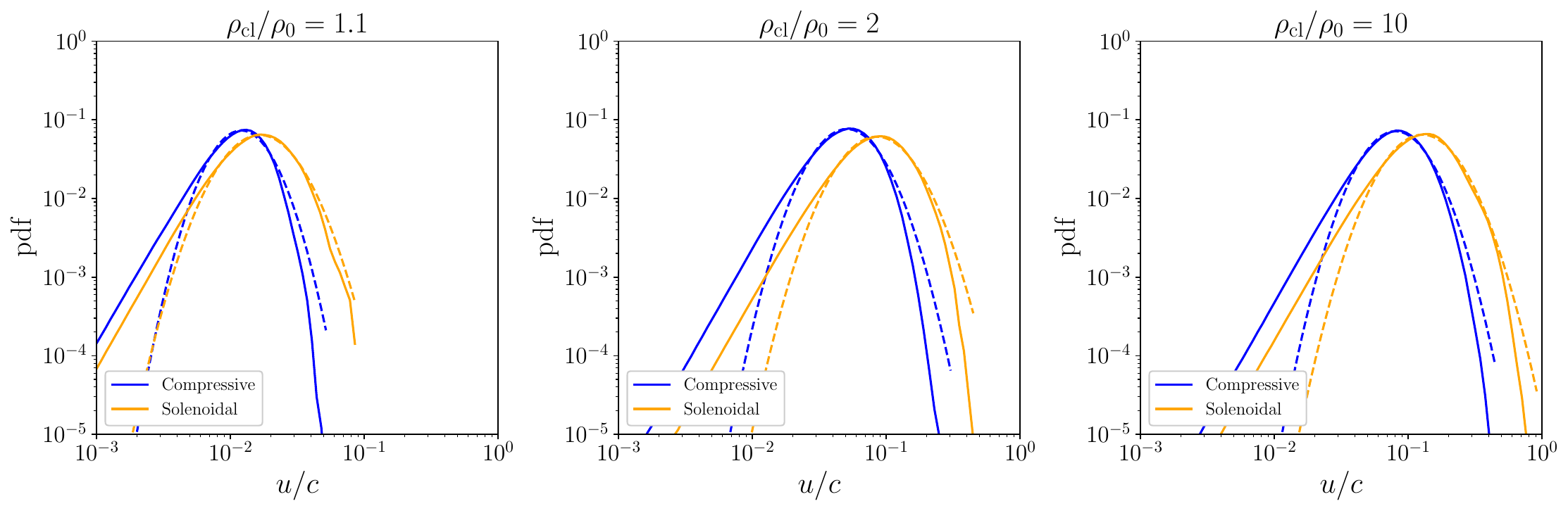}
   \caption{
   The same plot as Fig.~\ref{fig:pdf}, but for the absolute value of the compressive (blue lines) and solenoidal (orange lines) velocities. 
   The solid and dashed lines show the simulation results and log-normal functions fitted to the corresponding simulation data. 
   }
    \label{fig:pdf_decomposed}
\end{figure*}
Next, we plotted the probability distribution for the compressive and solenoidal velocities in Fig.~\ref{fig:pdf_decomposed}, where the left, middle, and right panels show the probability distributions for $\rho_{\rm cl}/\rho_0=1.1$, $2$, and $10$. 
The blue and orange solid lines show the probability distributions of compressive and solenoidal velocities in the simulation. 
The blue and orange dashed lines are log-normal functions fitted to the corresponding simulation
data.
It should be noted that these velocities are obtained in the simulation (the main shock rest) frame and the uniform velocity field is excluded. 
All the probability distributions have a similar shape although the peak location is different. 
In addition, the probability distributions only in the peaked region can be fitted by the log-normal function. 
As summarized in Tab.~\ref{tab:velocity_components}, the compressive mode has an amplitude comparable to that of the solenoidal mode, although the solenoidal mode has a larger amplitude.

\section{Discussion}
\label{sec:4}
In this work, we have investigated the downstream turbulence driven by the interaction of the relativistic shock with multiple clumps. 
The free energy for the downstream turbulence is the kinetic energy of the inhomogeneous upstream region in the downstream rest frame. 
As long as the total clump mass is sufficiently smaller than the total mass ($f \rho_{\rm cr}<\rho_0$), the free energy density in the downstream rest frame can be roughly estimated as $E_{\rm free} = f \Gamma_{\rm up,d}(\rho_{\rm cl,d}-\rho_{\rm 0,d}) c^2$, where $f, \Gamma_{\rm up,d}, \rho_{\rm cl,d}$ and $\rho_{\rm cl,d}$ are the volume filling factor, upstream Lorentz factor, the upstream uniform density and the upstream density in the clump in the downstream rest frame. 
If the free energy density is comparable to the downstream thermal energy density, $4\Gamma_{\rm up,d}\rho_{\rm 0,d}c^2$, we can expect strong turbulence in the downstream region. 
Since the volume filling factor is $f\approx 0.25$ in this work, strong turbulence is expected for $\rho_{\rm cl}/\rho_0=2$ and $10$, which is consistent with our simulation results. 

In reality, only a part of the free energy is converted to downstream turbulence. 
Some of the free energy is dissipated by the reflected shock and viscosity. 
The mean free path of the downstream particles could be larger than the size of the clump. 
Then, physics in collisionless plasmas could be important \citep{Ohira2016b,Ohira2016,Tomita2019,Tomita2022}.
In addition, if the energy density of the downstream magnetic field is comparable to the free energy density, the Lorentz force affects downstream turbulence \citep{Tomita2022}. 
Therefore, to understand the downstream turbulence and magnetic field amplification in the relativistic shock clump interaction, we have to investigate these topics further although understanding the collisionless kinetic effect is challenging.

Our simulations showed that there are secondary shocks and compressive modes in the shock downstream region.
Charged particles could be accelerated at the main relativistic shock by the 1st-order Fermi acceleration \citep{Axford1977, Krymskii1977, Blandford1978, Bell1978, Kirk2000}. 
In addition to the shock acceleration, we can expect other particle acceleration processes by the secondary shocks and compressive modes \citep{Yokoyama2020}.
If particles are reaccelerated by the secondary shocks, the energy spectrum of the accelerated particles is modified \citep{Melrose1993, Bykov1993, Inoue2010, Kang2021}. 
The secondary shock propagates to the downstream region of the main relativistic shock, in which the temperature is relativistically hot. 
Recently, collisionless shock formation and particle acceleration in a relativistically hot plasma have been demonstrated in particle-in-cell simulations \citep{Kamiido2024,Kamiido2025}. 
Thus, particles can be newly injected to the shock acceleration at each reflected shock, so that the number of non-thermal particles might increase with the downstream distance from the shock front. 
Even if particles are not injected to acceleration when they first interact with the main shock, if they are accelerated by secondary shocks or turbulence in the downstream region and go back to the main shock by diffusion, they could be further accelerated by the main shock \citep{Morikawa2024}. 
Therefore, the acceleration and the injection of non-thermal particles in the secondary shocks and downstream turbulence enhance the production of non-thermal particles.

Particles can be accelerated by the compressive and solenoidal turbulence \citep{Bykov1993, Ohira2013, Yokoyama2020} and the particle acceleration by turbulence in the shock downstream region has been widely discussed \citep{Liu2008, Pohl2015, Asano2015, Morikawa2024}. 
Particle acceleration by the collisionless kinetic turbulence is also investigated by Particle-in-Cell simulations \citep{Zhdankin2017, Comisso2018}. 
These studies often assume an isotropic turbulence or a solenoidal driving force for simplicity. 
However, our simulation showed that the downstream turbulence has compressive modes with an amplitude comparable to that of solenoidal modes for relativistic shocks. 
Moreover, the downstream turbulence is not isotropic. 
It should be studied in detail how particles are accelerated in the actual downstream turbulence driven by the relativistic shock-clump interaction.

In this work, the volume filling factor and the size of the density clumps were fixed. 
In reality, the density structure would have a power spectrum in the wave number space. 
Furthermore, the amplitude of inhomogeneity, the volume filling factor, and the power spectrum would depend on the environment of objects, which could be the origin of the wide distribution of $\epsilon_{\rm B}$ observed in GRB afterglows and the diversity of the spectrum of accelerated particles observed in high energy astronomical objects (GRB, GRB afterglow, PWNe, AGN, and so on). 
Understanding the relationship between the density structure and non-thermal emission is crucial for identifying the origin of high-energy astronomical phenomena.
Therefore, this work should be extended to cases of different density structures in order to achieve this goal.

\section{Summary}
\label{sec:5}
We have studied the downstream turbulence driven by the interaction of the relativistic shock with multiple clumps using relativistic MHD simulations. 
By performing the Helmholtz decomposition of the turbulent velocity field in the downstream flow, 
we found that, in contrast to the non-relativistic case, compressible modes have an amplitude comparable to that of solenoidal modes in the downstream turbulence. 
In addition, we found many reflected shocks and the anisotropy in the downstream turbulence. 
Thus, our simulation showed that the turbulence in the downstream region of relativistic shocks propagating in inhomogeneous media is not an isotropic incompressible turbulence, as is often assumed to study particle acceleration and magnetic field amplification by turbulence.  

The magnetic field is amplified in the shock downstream region by dynamo mechanism \citep{Kulsrud1992, Xu2016}. 
The ratios of the magnetic energy to the upstream kinetic energy, $\epsilon_{\rm B}=B^2/8\pi\bar{\rho}_{\rm up}(\Gamma_{\rm sh}-1)c^2$ is about $10^{-4}-10^{-3}$ for $\rho_{\rm cl}/\rho_0=2$, and $10$, but  $\epsilon_{\rm B}\sim 10^{-5}$ for $\rho_{\rm cl}/\rho_0=1.1$. 
The dependence of the upstream density inhomogeneity in this work suggests that the wide distribution of $\epsilon_{\rm B}$ in the observation of GRB afterglows indicates the diversity of the GRB environment. 

The probability distributions of the density and pressure are well described by the log-normal distribution, but for the magnetic field strength and the four-velocity, significant deviations from the log-normal distribution appear in their small values. 
To understand the intermittency more accurately, we need further studies with a higher numerical resolution.

\section*{Acknowledgements}
We are grateful to Yosuke Matsumoto for sharing the SRCANS+ code.
K. M. is supported by JST SPRING, Grant Number JPMJSP2108, and supported by the International Graduate Program for Excellence in Earth-Space Science (IGPEES), The University of Tokyo.
Y. O. is supported by JSPS KAKENHI Grants No. JP21H04487, and No. JP24H01805.
Numerical computations were carried out on Cray XC50 at Center for Computational Astrophysics, National Astronomical Observatory of Japan.

\section*{Data Availability}
The data underlying this article will be shared on reasonable request to the corresponding author.




\bibliographystyle{mnras}
\bibliography{00-reference} 






\bsp	
\label{lastpage}
\end{document}